\documentclass[11pt,notitlepage,tightenlines,nofootinbib,amsmath,amssymb,aps,prd]{revtex4-1}

\usepackage{graphicx}
\usepackage{subcaption}

\usepackage{amsmath,amssymb}
\usepackage{amsfonts,amssymb,mathrsfs}

\usepackage[bookmarks=false,pdfstartview=FitH]{hyperref}
\usepackage[all]{hypcap}

\def\be{\begin{equation}}
\def\ee{\end{equation}}

\def\f{\frac}

\def\pl{{\rm Pl}}
\def\lp{\ell_\pl}
\def\extd{{\rm d}}

\usepackage{color}

\begin{document}

\pagestyle{plain}

\title{Potential consequences of wormhole-mediated entanglement}

\author{Edward Wilson-Ewing} \email{edward.wilson-ewing@unb.ca}
\affiliation{Department of Mathematics and Statistics, University of New Brunswick, 
Fredericton, NB, Canada E3B 5A3}

\begin{abstract}

There are hints that the connectivity of space-time in quantum gravity could emerge from entanglement, and it has further been proposed that any two entangled particles may be connected by a quantum wormhole.  One way to test this proposal is by probing the electric field of an entangled charged particle to determine whether its electric field leaks through the putative wormhole.  In addition, if such a wormhole is traversable, then it could be possible for the collapse of the wave function to occur in a causal manner, with information about the collapse travelling through the wormhole at the speed of light, rather than the wave function collapse being a global and instantaneous event.

\end{abstract}

\maketitle

\section{Introduction}
\label{s.intro}

Recent work in quantum gravity suggests entanglement is closely related to space-time connectivity, and may in fact generate it.  This was first noticed in the context of the duality between asymptotically anti-de Sitter space-times and conformal field theories (AdS/CFT), where the connectivity of the space-time geometry in the bulk AdS theory is closely related to entanglement entropy in the dual CFT \cite{Ryu:2006bv, Ryu:2006ef, VanRaamsdonk:2010pw, Lashkari:2013koa, Faulkner:2013ica}.

These results, derived in the context of AdS/CFT, have since been extended to suggest that the microscopic connectivity of space-time may be closely related to (or perhaps even determined or caused by) entanglement entropy in \emph{any} quantum gravity theory \cite{Bianchi:2012ev, Swingle:2014uza}.  In addition to AdS/CFT, this also turns out to be the case in, for example, loop quantum gravity where the connectivity between neighbouring quanta of geometry can be generated by entangling the quantum degrees of freedom that live on the surface areas that are to be glued together \cite{Donnelly:2008vx, Oriti:2013aqa, Chirco:2017vhs, Baytas:2018wjd}.  Thus, this body of work in various approaches to quantum gravity suggests that the connectivity of an emergent space-time may be determined by the entanglement of geometric quanta (or their CFT duals).

According to these ideas, if there is some entanglement between geometric degrees of freedom lying in two otherwise disparate regions, this entanglement will generate a quantum gravity wormhole linking these two regions.  Depending on the specific theory of quantum gravity, the wormhole may be represented in different ways---in background-independent approaches to quantum gravity the connectivity information is often captured combinatorially, for example as links between nodes on an abstract graph in which case a `wormhole' would simply be a link connecting two otherwise distant nodes in the graph.  In other words, the fundamental notion of locality (based on combinatorial microscopic degrees of freedom) and a coarse-grained notion of locality (based on an emergent semi-classical space-time geometry) may not always agree  \cite{Smolin:1994uz, Markopoulou:2007ha, Konopka:2008hp, PrescodWeinstein:2009wq, Hossenfelder:2013yda}.

In an additional step, it has been further conjectured that any two particles (and not only quanta corresponding to the gravitational field or its conformal dual) that are entangled must be connected by a wormhole \cite{Maldacena:2013xja}; for example, this would be the case for two electrons whose spins are entangled.  This proposal is also known as ER=EPR, with the name suggesting that Einstein-Podolsky-Rosen entangled pairs are connected by microscopic Einstein-Rosen bridges, i.e., wormholes.  Note that the wormhole between two entangled particles will collapse if a measurement destroys the entanglement.

Based on the AdS/CFT mapping between an eternal two-sided black hole (that has a space-like wormhole connecting the two different asymptotic regions) and the thermofield double entangled state for the two boundary CFTs, it was originally suggested that the geometry of ER=EPR wormholes connecting entangled particles might be analogous to the non-traversable wormholes that arise in the eternal (AdS) Schwarzschild black hole space-time.  However, it was soon pointed out that the properties of some forms of tripartite entanglement like the 3-particle Greenberger-Horne-Zeilinger (GHZ) state $|\psi\rangle = (|000\rangle + |111\rangle)/\sqrt2$ \cite{Greenberger:1989} cannot be captured by a classical geometry \cite{Gharibyan:2013aha, Balasubramanian:2014hda}, ruling out the possibility that these `entanglement' wormholes could potentially be described in a classical fashion.  As a result, if the ER=EPR conjecture is correct and entanglement does in fact generate wormholes, it seems that these wormholes must be treated as fully quantum objects and it is not possible (perhaps even for the relatively simple case of bipartite entanglement) to describe their geometry classically.  Further, it has since also been pointed out that traversable wormholes can be created without needing exotic matter fields, either by adding interactions (for wormholes between two-sided black holes) \cite{Gao:2016bin} or by including fermionic matter \cite{Blazquez-Salcedo:2020czn}; in both cases causal signals can travel from one side of the wormhole to the other.  And from the perspective of background-independent approaches to quantum gravity with combinatorial microscopic degrees of freedom, the simplest quantum wormholes (composed for example of a single link connecting otherwise distant nodes in a graph) would also be traversable.  These considerations, combined with the ER=EPR conjecture, suggests that if two interacting particles are entangled then the wormhole conjectured to connect them might very well be traversable.

In the following, I explore two possible consequences of the ER=EPR conjecture for entangled particles, allowing for the possibility that the wormhole may be traversable or not.  For the sake of simplicity I will only consider systems of two entangled particles, and leave the study of systems with three or more entangled particles, like GHZ states, for future work.

First, one way to test the ER=EPR conjecture is to measure the electric field surrounding an entangled charged particle.  In the presence of a wormhole, part of the electric field sourced by the particle would leak into the wormhole \cite{Dai:2019mse} and this effect would change the strength of the electric field surrounding an entangled charged particle, as compared to the unentangled case.  (For other possible tests of the ER=EPR conjecture, see \cite{Dai:2020ffw}.)

Second, if quantum wormholes arise in the presence of entanglement as suggested by the ER=EPR conjecture, then the microscopic structure of space-time may play an important role in explaining at least some aspects of quantum mechanics.  Interestingly, this is in broad agreement with earlier suggestions that it may be necessary not only to `quantize gravity' but also to `gravitize quantum mechanics' \cite{Penrose:2014nha}, with the ER=EPR conjecture suggesting that perhaps a key part of `gravitizing' quantum mechanics is to include effects from the non-trivial topology of the microscopic structure of space-time for entangled states.  If this is indeed the case, it may be impossible to fully understand quantum mechanics without including gravity \cite{Susskind:2016jjb}.  In particular, the apparently instantaneous collapse of the wave function seems to violate the usual laws of locality and causality in physics.  But if there is a traversable wormhole connecting entangled particles, then this wormhole would provide the necessary shortcut where the information about the collapse of the wave function could travel at finite speed after a local measurement is performed on one of the entangled particles, even though the collapse of the wave function would appear to occur essentially instantaneously according to an outside observer unaware of the wormhole.

\section{A Potential Observational Test}
\label{s.test}

If the ER=EPR conjecture is correct and two entangled particles are joined by a wormhole, it seems reasonable to expect that a sufficiently weak electromagnetic field (namely, a field that does not affect the entanglement between the two particles) can be expected to enter the wormhole \cite{Dai:2019mse}, without causing the wormhole to collapse.  (And if the wormhole is traversable, then these fields would eventually spill out near the particle's entangled pair.)  If this is indeed the case, then it is possible to test the ER=EPR proposal.

Consider two charged particles that are entangled with each other, and for simplicity assume that they are stationary and distant from each other.  In particular, the electric fields generated by these two charged particles will leak into the quantum wormhole connecting them, so one way to test whether a wormhole is present is to measure whether the electric field around either of the entangled particles is different than for a non-entangled particle: is the electric field leaking into the wormhole?

There are two main possibilities to consider: either the wormhole is traversable or not.  If the wormhole is non-traversable, then the only effect would be that some of the electric field would leak into the wormhole, weakening the electric field surrounding each of the entangled particles.  On the other hand, if the wormhole is traversable there would be the additional effect of the electric field due to its entangled pair leaking out through the wormhole.

To make this discussion concrete, I will consider two stationary particles $A$ and $B$ with electric charges of $q_A$ and $q_B$ that are entangled and separated by a large distance, and I will focus on the electric field near particle $A$.  Since the particles are separated by a large distance, in the absence of a wormhole the electric fields near particle $A$ would be determined to a very good approximation entirely by $q_A$ and the distance from particle $A$ by Coulomb's law, with contributions from particle $B$ being negligible.

If some of the electric field produced by particle $A$ leaks into the wormhole, then this effect will weaken the electric field surrounding particle $A$.  By Gauss's law, the charge of particle $A$ is given by the surface integral of the electric field for a surface that encloses particle $A$ only.  And if there is a wormhole then it is necessary to include an extra contribution due to the electric field passing through the wormhole.  For this stationary configuration, in units where $\epsilon_0=1$,
\be
q_A = \oint_S \vec E_S \cdot \extd \vec a + \oint_W \vec E_W \cdot \extd \vec a,
\ee
where $\oint_W$ denotes the surface integral that is transverse to the wormhole, and $\oint_S$ denotes the surface integral on a sphere $S$ which is centred at the particle $A$ and has a radius larger than the distance of the mouth of the wormhole from particle $A$.

Alternatively, if particle $A$ is not entangled there is no wormhole and, choosing the same $S$,
\be
q_A = \oint_S \vec E_S' \cdot \extd \vec a.
\ee
It is clear that $|\vec E_S'| \neq |\vec E_S|$ if $\vec E_W \neq 0$: the electric field surrounding particle $A$ changes depending on whether particle $A$ is entangled or not.

If the wormhole is non-traversable, the only contribution to $\vec E_W$ comes from particle $A$.  In this case $|\vec E_S'| > |\vec E_S|$ and the electric field surrounding particle $A$ will be weaker if particle $A$ is entangled.

On the other hand, if the wormhole is traversable then there will be another contribution to $\vec E_W$ from particle $B$.  One possibility is that particle $B$ has the same charge, $q_A = q_B$, and then by symmetry $\vec E_W = 0$ in which case there will be no effect due to the presence of the wormhole.  But if $q_A \neq q_B$, then $\vec E_W \neq 0$ and the electric field around particle $A$ will depend on whether it is entangled or not: $|\vec E_S'| \neq |\vec E_S|$.  In particular, if particle $B$ has the opposite charge of particle $A$, then this effect will be twice as strong compared to the case of a non-traversable wormhole.

Note that there is one possible way out, in which case there is no effect at all: if each unentangled particle is attached to a `bridge to nowhere' wormhole as suggested (for black holes) in \cite{Susskind:2014jwa}, while entangled particles are joined by a non-traversable wormhole.  In this case, the same electric field $\vec E_W$ would be lost whether the particle is entangled (in this case, the electric field $\vec E_W$ leaks into the non-traversable wormhole) or not (in this case, the same $\vec E_W$ leaks into the `bridge to nowhere' wormhole).  But if the wormhole is traversable, or if an unentangled particle is not connected to a wormhole, then the electric field around a charged particle will depend on whether that particle is entangled or not.

In the absence of a complete theory of quantum gravity, it is impossible to know the precise properties of these conjectured `entangling' quantum wormholes and accurately predict the strength of this effect on the electric field.  However, heuristic arguments can provide an order of magnitude estimate.  Due to arguments based on applying the ER=EPR conjecture to a pair of entangled black holes, it has been conjectured that the total transverse area $a_t$ of a wormhole is related to the entanglement entropy by the relation $a_t = 4 s \lp^2$, with $s$ the entanglement entropy between particles $A$ and $B$ \cite{Verlinde:2020upt}; for example, for two electrons whose spins are maximally entangled $s = \ln 2 \sim 0.69$.  Further, the mouth of the wormhole could reasonably be expected to be a distance $\delta \sim \alpha \lp$ from particle $A$, with $\alpha$ a dimensionless constant of order unity.

Considering the case of a non-traversable wormhole first, only the electric field due to particle $A$ will contribute to $\vec E_W$.  I will assume that the electric field entering the wormhole due to particle $A$, whose mouth is a distance $\delta$ from the particle, has the same norm $|\vec E_W|$ as the electric field at any point on the sphere $S_\delta$ with the same radius $\delta$, so $|\vec E_W| = |\vec E_{S_\delta}|$.  With $\oint_{S_\delta} \vec E_{S_\delta} \cdot d \vec a = 4 \pi \alpha^2 \lp^2 |\vec E_{S_\delta}|$ and $\oint_W \vec E_W \cdot d \vec a = a_t |\vec E_W| = 4s \lp^2 |\vec E_W|$, it follows that
\be
q_A = (4 \pi \alpha^2 + 4s) \lp^2 |\vec E_{S_\delta}|,
\ee
implying that $|\vec E_{S_\delta}| = |\vec E_W| = q_A/(4 \pi \alpha^2 + 4s)\lp^2$.  Knowing $|\vec E_W|$, it is now easy to determine the strength of the electric field $\vec E_S$ at a distance $r > \delta$ from particle $A$, with the expected result that it decreases as $|\vec E_S| = (\delta^2/r^2) |\vec E_{S_\delta}|$.  Compared to an unentangled particle, the electric field at a distance $r$ of an entangled particle is
\be \label{e-ntr}
\vec E_S = \f{\vec E_S'}{1+s/\pi\alpha^2},
\ee
where the electric field surrounding an unentangled particle of charge $q_A$ is given by the usual Coulomb law, $\vec E_S' = (q_A/4\pi r^2) \hat r$.

For a traversable wormhole, given the same assumptions,
\be
\oint_W \vec E_W \cdot d \vec a = \f{s (q_A-q_B)}{\pi \alpha^2 + s},
\ee
where there are now contributions from both particles $A$ and $B$ to $\vec E_W$.  Then,
\be
|\vec E_S| = \f{1}{4 \pi r^2} \left( q_A - \f{s (q_A-q_B)}{\pi \alpha^2 + s} \right),
\ee
or, as compared to the electric field surrounding an unentangled particle,
\be \label{e-tr}
\vec E_S = \left( 1 - \f{(q_A-q_B)}{q_A} \f{s}{\pi \alpha^2 + s} \right) \vec E'_S.
\ee

Based on these estimates, Eq.~\eqref{e-ntr} suggests that for a non-traversable wormhole, if $\alpha \sim 1$ this is a $\sim 25\%$ effect while if $\alpha \sim 5$ there is a $\sim 1\%$ effect, in both cases assuming $s \sim 1$.  For traversable wormholes, the amplitude of the effect depends on the relative charges $q_A$ and $q_B$; as compared to non-traversable wormholes the effect will be stronger in the case of traversable wormholes if the entangled particles have charges of opposite sign, but weaker if $q_A$ and $q_B$ have the same sign (and the weakened electric field surrounding particle $A$ will be the same if $q_B=0$ whether the wormhole is traversable or not).  For example, if $q_A$ and $q_B$ are equal and opposite, and again taking $s \sim 1$, then Eq.~\eqref{e-tr} indicates that for traversable wormholes if $\alpha \sim 1$ this is a $\sim 50\%$ effect while if $\alpha \sim 5$ there is a $\sim 2\%$ effect.  On the other hand, if $q_A=q_B$ then there is no modification to the electric field (as compared to the unentangled case) surrounding either particle for such an EPR pair connected by a traversable wormhole.

Finally, note that this effect is a modification of the electric field surrounding an entangled charged particle.  But the charge of the particle remains the same, so it will respond in the same way to an external electromagnetic field whether it is entangled or not.

\section{Causal Collapse of the Wave Function}
\label{s.collapse}

An important open question in the foundations of quantum mechanics is the following, expressed in simplistic terms: if two particles are entangled, and a measurement is made on one of these particles, how does the other particle `know' that its wave function must collapse also?  According to the Copenhagen interpretation the entire wave function collapses instantaneously, but this is not fully satisfactory from the perspective of special relativity where the notion of `instantaneously' is frame-dependent.  Nonetheless, despite this theoretical prejudice, experiments clearly show that the collapse occurs much more rapidly than what would be possible if information about a local measurement were transmitted at the speed of light (or slower) \cite{Aspect:1982, Weihs:1998}.

Although the collapse of a wave function cannot be used to transmit information faster than the speed of light \cite{Ghirardi:1980sh, Eberhard:1988yj}, it remains somewhat unsettling that the wave function of a distant entangled particle immediately collapses when its entangled pair is measured.  However, this apparent tension can be resolved if there exists a short traversable wormhole that lies between two entangled particles through which the information can travel.

There are some indications that this might indeed be the case according to the ER=EPR conjecture.  In the context of quantum teleportation, information about a quantum state can be split into quantum EPR correlations and classical information \cite{Bennett:1992tv}, and at least for eternal black holes the quantum correlations in a quantum teleportation experiment appear to travel through the wormhole after a measurement \cite{Marolf:2012xe, Numasawa:2016emc, Susskind:2016jjb}.  For the problem of the collapse of the wave function, this quantum information---passing through the wormhole---is precisely what is needed to update the wave function of an entangled particle after its EPR pair has been measured.  (Of course, no additional classical information is needed in this case.)

If the wormhole is traversable, it offers a shortcut that makes a causal collapse of the wave function possible: even if two entangled particles appear to be far away, information can pass through the wormhole.  Assuming the information is transmitted at the speed of light and the length of the wormhole is $\ell$, it will only take the time $\Delta t = \ell/c$ for information to travel from one entangled particle to the other.

If the wormhole is sufficiently short compared to the distance $d$ between the two entangled particles as measured by an outside observer who does not know that the wormhole is there, $\ell \ll d$, then any signals passing through the wormhole will appear to have propagated at a superluminal speed according to outside observers unaware of the wormhole (while, of course, the signal travelled at the speed of light through the wormhole).  In particular, if a measurement is performed on one of the particles, the wormhole provides a causal mechanism by which the information (i) that a measurement was made and (ii) of its outcome can be sent to its \mbox{(ex-)entangled} partner in a nearly-instantaneous fashion that nonetheless only travels at the speed of light.  Of course, this signal must pass through the wormhole before it collapses due to the measurement.

In short, the wormhole gives a non-trivial topology to space-time which can resolve the tension between the speed limit of $c$ for any signal and the apparent instantaneous collapse of the wave function.  (Also, the presence of a wormhole provides, at least in principle, the non-locality necessary so that the locality assumption underlying Bell's theorem \cite{Bell:1964} does not hold, and could possibly open the way to constructing a hidden variable theory for quantum mechanics; see \cite{Markopoulou:2003ps} for one such proposal.)

Finally, the presence of a traversable wormhole raises the possibility of using it to send messages in an apparently superluminal fashion.  Of course, any information sent through the wormhole is moving at most at the speed of light, but to observers unaware of the existence of the wormhole this signal seems to be transferred from one entangled partner to the other superluminally.  For this to be possible, the wormholes must be traversable, sufficiently robust to allow at least a weak signal to pass through them, and the signal must not affect the entanglement between the particles---or otherwise the wormhole would collapse, destroying the shortcut.  This would be difficult since the mouth of the wormhole would presumably be very close to the entangled particle, so it would likely be hard to send signals into the wormhole without disturbing the particle and collapsing the wave function and the wormhole with it.  In fact, the entangled particle at the mouth of the wormhole may block such signals entirely.  But if not, this would open the possibility of sending signals in an essentially instantaneous fashion between distant laboratories that share entangled particles.

(If it is possible to use these wormholes to transmit information, then it may be possible at least in principle to form closed time-like curves by accelerating one of the entangled particles, and hence one of the mouths of the wormhole \cite{Morris:1988tu}, so long as the entanglement is preserved.  Also, there may be a bound on the amount of information that can be transferred through the wormhole \cite{Maldacena:2017axo}, but this bound alone does not rule out sending signals with limited information.)

Of course, the discussion in this section has assumed not only that there is a wormhole connecting two entangled particles, but also that this wormhole is traversable.  It is clear that even if there is a wormhole but it is non-traversable, then it cannot provide a mechanism for a causal collapse of the wave function.

\section{Discussion}
\label{s.disc}

The ER=EPR conjecture proposes that a pair of entangled particles are connected by a quantum wormhole \cite{Maldacena:2013xja}.  A possible test of this proposal would be to measure the electric field around entangled and unentangled charged particles; different results would be expected if this conjecture is correct, due to the electric field leaking into the wormhole when a particle is entangled.

If this effect were to be detected, its amplitude would determine whether the wormhole is traversable or not, since if the wormhole is traversable then the electric field around an isolated entangled particle also depends on the charge of its entangled pair.  Note that a challenge from the experimental perspective is that this requires a precise measurement of the electric field close to an entangled particle, but without causing the wave function to collapse.

Finally, if the ER=EPR conjecture is correct, and if in addition the wormholes connecting entangled particles are traversable, this would open the possibility for a causal collapse of the wave function after a measurement: information could pass through the wormhole in a way that appears to be instantaneous for an outside observer unaware of the existence of the wormhole.

\bigskip

\noindent
{\it Acknowledgments:} 
This work was supported in part by the Natural Sciences and Engineering Research Council of Canada and the UNB Fritz Grein Award.

\newpage

\small

\raggedright

\end{document}